\documentclass[aps,pre,twocolumn,superscriptaddress,superscriptreference]{revtex4-1}

\usepackage{amsmath,bbold,bm,amssymb,scalerel,mathtools}
\usepackage{graphicx}
\usepackage{color}
\usepackage{enumitem}
\usepackage{algorithm,algpseudocode}
\usepackage{multirow}
\usepackage{colortbl,booktabs}
\usepackage{placeins}
\usepackage[usenames,dvipsnames]{xcolor}

\usepackage[colorlinks, linkcolor=BrickRed, urlcolor=blue!50!black, citecolor=blue!50!black]{hyperref}

\newcommand*{\citen}{}
\DeclareRobustCommand*{\citen}[1]{%
	\begingroup
	\romannumeral-`\x 
	\setcitestyle{numbers}%
	\cite{#1}%
	\endgroup
}

\newcommand\equalhat{%
	\let\savearraystretch\arraystretch
	\renewcommand\arraystretch{0.3}
	\begin{array}{c}
		\stretchto{
			\scalerel*[\widthof{=}]{\wedge}
			{\rule{1ex}{3ex}}%
		}{0.5ex}\\ 
		=%
	\end{array}
	\let\arraystretch\savearraystretch
}
\newcommand{\myrowcolour}{\rowcolor[gray]{0.925}}

\newcommand\diff{\mathrm{d}}

\newcommand{\gj}[1]{\textcolor{black}{#1}}

\usepackage[normalem]{ulem}

\begin{document}

\title{\gj{Dynamical properties of densely packed confined hard-sphere fluids}}

\author{Gerhard Jung}
\email{gerhard.jung@uibk.ac.at}
\affiliation{Institut f\"ur Theoretische Physik, Universit\"at Innsbruck, Technikerstra{\ss}e 21A, A-6020 Innsbruck, Austria}

\author{Michele Caraglio}
\affiliation{Institut f\"ur Theoretische Physik, Universit\"at Innsbruck, Technikerstra{\ss}e 21A, A-6020 Innsbruck, Austria}

\author{Lukas Schrack}
\affiliation{Institut f\"ur Theoretische Physik, Universit\"at Innsbruck, Technikerstra{\ss}e 21A, A-6020 Innsbruck, Austria}

\author{Thomas Franosch}
\affiliation{Institut f\"ur Theoretische Physik, Universit\"at Innsbruck, Technikerstra{\ss}e 21A, A-6020 Innsbruck, Austria}

\begin{abstract}
Numerical solutions of the mode-coupling theory (MCT) equations for a hard-sphere fluid confined between two parallel hard walls are elaborated. The governing equations feature multiple parallel relaxation channels which significantly complicate their numerical integration. We investigate the intermediate scattering functions and the susceptibility spectra close to structural arrest and compare to an asymptotic analysis of the MCT equations. 
We corroborate that the data converge in the $\beta$-scaling regime to two asymptotic power laws, viz. the critical decay and the von Schweidler law. The numerical results reveal a non-monotonic dependence of the power-law exponents on the slab width and a non-trivial kink in the low-frequency susceptibility spectra. We also find  qualitative agreement of these theoretical results to event-driven molecular-dynamics simulations of polydisperse hard-sphere system. In particular, the non-trivial dependence of the dynamical properties on the slab width is well reproduced.

\end{abstract}

\maketitle

\section{Introduction}

The structural relaxation of dense liquids displays a drastic slowing down of transport upon compression or cooling, a phenomenon commonly identified with the glass transition~\cite{Gotze2009}. 
Upon approaching this transition, a supercooled bulk liquid exhibits several fascinating dynamical properties, in particular, the aforementioned slowing down, stretching of the intermediate scattering function, as well as a two-step power-law-relaxation behavior. 
All of these features have been observed in experiments \cite{VanMegen1993,VanMegen1994,Li:PRA_45_1992,Li:PRA_46_1992,Franosch:PRE_56_1997,Singh1998,Lunkenheimer:PRL_77_1996,Schneider1999,Goetze2000} and simulations \cite{Kob1994,Horbach2008,Gallo:PRL_76_1996,Sciortino:PRE_54_1996,Sciortino:PRE_56_1997,Sciortino:PRL_86_2001,Horbach1998,Voigtmann2006} and were successfully described by mode-coupling theory of the glass transition (MCT)~\cite{Bengtzelius_1984,Gotze_1992,Goetze:JPC:1999,Gotze2009,JansenReview2018}.
The underlying microscopic picture behind MCT is the trapping of particles in transient ``cages'' which are formed by their respective neighbors.
Consequently, confining the supercooled liquid between two parallel walls is expected to have a significant impact on the nature of the glass transition~\cite{Alba_Simionesco:JPC:2006,Lang2010,Varnik:JPC:2016}.
Strong confinement will hinder or promote the formation of ``cages'' and the wall-particle interaction will induce layering~\cite{Nemeth:PRE59:1999,Mittal:PRL100:2008}, thus drastically changing the local structure in the fluid.
Analyzing the consequences of confinement is therefore of fundamental interest, not only to question the microscopic picture of the glass transition but also to better understand physical, chemical, and biological systems where confinement occurs naturally like porous rocks or crowded living cells.

In the last twenty years confined liquids have therefore been  studied extensively in
 experiments~\cite{C1SM06502E,PhysRevE.83.030502,PhysRevLett.99.025702,PhysRevLett.112.218302,doi:10.1063/1.4905472,PhysRevLett.108.037802,doi:10.1063/1.4825176,PhysRevLett.116.167801,PhysRevLett.117.036101,PhysRevLett.116.098302,PhysRevX.6.011014} and simulations
 \cite{Scheidler_2000,Scheidler_2002,doi:10.1021/jp036593s,PhysRevE.65.021507,PhysRevLett.85.3221,Baschnagel_2005,PhysRevLett.96.177804,doi:10.1063/1.2795699,doi:10.1021/jp071369e,Mittal:PRL100:2008,doi:10.1063/1.4959942,PhysRevLett.100.106001,doi:10.1063/1.3651478,PhysRevLett.111.235901,doi:10.1063/1.1524191,PhysRevE.86.011504,C3SM52441H,Mandal2017a}, showing that even general questions like whether confinement accelerates or hinders
 the dynamics depends on the nature of the wall-particle potential \cite{PhysRevLett.111.235901,C3SM52441H,PhysRevLett.100.106001} or the surface roughness \cite{Scheidler_2002,Baschnagel_2005,doi:10.1063/1.3651478}. Additionally, it has been found that confinement even with suppressed layering leads to significant changes in the static and dynamical properties of the liquid \cite{Scheidler_2000,Scheidler2000a,Petersen_2019,Schrack:JStatMech:2020}.

To better understand supercooled liquids in confinement MCT was extended to describe a simple fluid confined between two parallel, flat and hard walls (cMCT)~\cite{Lang2010, Lang2012,Mandal2014, Mandal2017a}.
The theoretical description depends on the introduction of symmetry-adapted fluctuating density modes which mirror the broken translational symmetry in lateral direction and lead to matrix-valued structure factors and intermediate scattering functions.
The validity of this description for the statical properties and nonergodicity parameters of hard-sphere glasses has been investigated and confirmed by event-driven molecular dynamics simulations \cite{Mandal2014, Mandal2017a}.
A remarkable feature of cMCT is the emergence of a multiple reentrant glass transition for liquids with constant packing fraction in a slab of variable slit width \cite{Lang2010, Lang2012}, which could also be observed using computer simulations \cite{Mandal2014, Mandal2017a}.
Interestingly, it also has been shown that the lines of constant chemical potential do not overlap with the glass-transition lines in a nonequilibrium state diagram, indicating that in a wedge geometry the observation of a coexistence between glass and liquid regions is anticipated.

A fundamental difference between the mathematical structure of the  MCT equations in confined geometry to MCT in simple liquids is the emergence of multiple relaxation channels.
These arise naturally due to a splitting of the particle-conservation law into distinct currents in lateral and transverse direction.
Recently, it has been shown for MCT with multiple relaxation channels that under very mild assumptions the $\beta$-scaling equation is still valid despite this change of structure \cite{Jung:JStatMech:2020}.
This allows us to perform a full asymptotic analysis of the dynamics by computing  the power-law exponents for the critical decay and the von Schweidler law from microscopic expressions.

Here we focus on the dynamical properties of strongly confined liquids as encoded in MCT. 
For the first time we present a numerical solution for the full time dependence of the cMCT equations. We investigate the behavior of the 
dynamic correlation functions close to structural arrest and compare 
the results to glass-forming liquids in bulk. Additionally, we  calculate the asymptotic power laws using the analysis presented in Ref.~\citen{Jung:JStatMech:2020} and show that they accurately describe the numerical 
solution in the $ \beta $-scaling regime. We  also investigate similarities and  differences of our theoretical 
results and  event-driven molecular dynamics simulations. The overall goal is to gain a better insight into the effects of confinement on supercooled liquids.

The manuscript is organized as follows: In Sec.~\ref{sec:mct_slab} we  recapitulate the mode-coupling theory in slit geometry and present 
the equations of motion for the intermediate scattering function. 
In order to enable a numerical integration we reformulate these equations by introducing an effective memory and apply an additional diagonal approximation. Afterwards, the numerical solution of the cMCT equations for the intermediate scattering function is presented in Sec.~\ref{sec:asymptotic_analysis} and compared to the asymptotic analysis. In Sec.~\ref{sec:comparison_edmd} we  then compare the dynamics predicted by cMCT to computer simulations. We summarize and conclude in Sec.~\ref{sec:conclusion}.

\section{Mode-coupling theory in slab geometry}
\label{sec:mct_slab}

\subsection{Equations of motion}

The fluctuating density modes of $ N $ particles confined in a channel of accessible width $ L $ can be introduced as \cite{Lang2010},
\begin{equation}\label{key}
\rho_\mu(\bm{q},t) = \sum_{n=1}^{N} \exp\left[ {\rm i} Q_\mu z_n(t) \right] e^{ {\rm i} \bm{q}\cdot \bm{r}_n(t)},
\end{equation}
with particle positions $ \bm{x}_n=(\bm{r}_n,z_n) $, restricted to the positions $ -L/2 \leq z_n \leq L/2 $, wave vectors $ \bm{q}=(q_x,q_y) $ and wavenumbers $ Q_\mu = 2 \pi \mu/L. $ In the following, 
we will refer to the indices $ \mu \in \mathbb{Z} $ as \emph{mode indices}. The packing fraction is defined as $ \varphi = N\pi \sigma^3/6V $, with particle diameter $ \sigma $, 
volume $ V = A H $, wall area $ A$ and wall distance $ H = L +\sigma $. Directly connected to the fluctuating density modes is its coherent time-dependent correlation function,
\begin{equation}\label{key}
S_{\mu\nu}(q,t)  = \frac{1}{N} \left \langle  \rho_\mu(\bm{q},t)^* \rho_\nu(\bm{q},0) \right \rangle.
\end{equation}
Its initial value $ S_{\mu\nu}(q) =  S_{\mu\nu}(q,t=0)$ is the structure factor, generalized to the slit geometry.

For this setting the Zwanzig-Mori projection operator formalism~\cite{Zwanzig2001,Gotze2009,Hansen:Theory_of_Simple_Liquids} with $ \left\{ \rho_\mu(\bm{q},t) \right\} $ as set of distinguished variables was applied \cite{Lang2010,Lang2012} to derive the equations of motion for the intermediate scattering functions $ S_{\mu\nu}(q,t) $,
\begin{equation}\label{eq:eom1}
\dot{\mathbf{S}}(t)+\int_0^t \mathbf{K}(t-t')\mathbf{S}^{-1}\mathbf{S}(t') \diff t' =0.
\end{equation}
Here, we have dropped the explicit dependence on the (magnitude of the) wave vectors $q$ and introduced the matrix notation $ \left[ \mathbf{S}(t) \right]_{\mu \nu} = S_{\mu\nu}(q,t) $.
The memory kernel $ \mathbf{K}(t) $ describes the non-Markovian dynamics of the intermediate scattering function and is related to the dynamic correlation function of the density modes, $  \dot{\rho}_\mu(\bm{q},t) $.

To derive an expression for the memory kernel, we consider the continuity equation for the density modes,
\begin{equation}\label{key}
\dot{\rho}_\mu(\bm{q},t) = {\rm i} \sum_{\alpha =\parallel,\perp }b^\alpha (q,Q_\mu) j_\mu^\alpha(\bm{q},t),
\end{equation}
with the selector $ b^\alpha(x,z) = x \delta_{\alpha \parallel} + z \delta_{\alpha \perp}  $ depending on the \emph{channel index} $\alpha \in \{ \parallel, \perp\}$, and the current channels $ j^\alpha_\mu(\bm{q},t) $. 
Since the currents in lateral ($ \perp $) and longitudinal ($\parallel$) direction are anticipated to behave differently, the memory kernel $ \mathbf{K}(t) $ splits naturally into multiple decay channels, 
\begin{align}\label{key}
K_{\mu\nu}(q,t)&= \left[ \mathcal{C} \{\bm{\mathcal{K}} \}  \right]_{\mu \nu} \nonumber \\
&\coloneqq 
\sum_{\alpha,\beta=\parallel,\perp} b^\alpha (q,Q_\mu) \mathcal{K}^{\alpha \beta}_{\mu \nu} (q,t) b^\beta(q,Q_\nu).
\end{align}
 By performing a second Zwanzig-Mori projection step using the current modes $\{j_\mu^\alpha(\bm{q},t)\}$ as distinguished variables the equations of motion for the components of the memory kernel $ \mathcal{K}^{\alpha \beta}_{\mu \nu}(q,t) $ can be derived. This yields the result,
\begin{equation}\label{eq:eom2}
\bm{\mathcal{J}}^{-1} \dot{\bm{\mathcal{K}}}(t) + \bm{\mathcal{D}}^{-1}\bm{\mathcal{K}}(t) + \int_0^t \bm{\mathcal{M}}(t-t')\bm{\mathcal{K}}(t') \diff t'= 0,
\end{equation}
with the matrix notation $ \left[ \bm{\mathcal{K}}(t) \right]_{\mu\nu}^{\alpha \beta} = \mathcal{K}_{\mu\nu}^{\alpha \beta}(q,t)  $ and  
\begin{equation}\label{key}
\mathcal{J}^{\alpha\beta}_{\mu\nu}(q) = \mathcal{K}^{\alpha\beta}_{\mu\nu}(q,t=0) = v_{\text{th}}^2 \frac{n_{\mu - \nu}^*}{n_0} \delta_{\alpha \beta}.
\end{equation}
Here, we have introduced the thermal velocity $v_{\text{th}} = \sqrt{k_B T/m}$ related to particle mass $m$ and the thermal energy $ k_\text{B} T$, as well as the decomposition of the inhomogeneous density profile $ n(z) $ into  Fourier modes 
\begin{equation}\label{key}
n_\mu = \int_{-L/2}^{L/2} n(z) \exp \left[ {\rm i} Q_\mu z \right] \diff z, \qquad \mu\in \mathbb{Z}.
\end{equation}
We also included an instantaneous damping term given by a positive semidefinite Hermitian matrix in the mode and channel  indices, $ \bm{\mathcal{D}}^{-1} \succeq 0$.

Mode-coupling theory now provides an approximation for the force kernel $ \bm{\mathcal{M}}(t) $ as a bilinear functional of the intermediate scattering functions  $ S_{\mu\nu}(q,t) $~\cite{Lang2010,Lang2012},
\begin{equation}\label{key}
\mathcal{M}_{\mu \nu}^{\alpha \beta}(q,t) = \mathcal{F}^{\alpha \beta}_{\mu \nu} \left[ \mathbf{S}(t);q \right],
\end{equation}
with,
\begin{align}\label{eq:MCT_functional}
\mathcal{F}^{\alpha \beta}_{\mu \nu} \left[\mathbf{S}(t);q \right] &= \frac{1}{2N} \sum_{\substack{\bm{q}_1,\\\bm{q}_2=\bm{q}-\bm{q}_1}} \sum_{\substack{\mu_1,\mu_2\\\nu_1,\nu_2}} \mathcal{Y}^\alpha_{\mu \mu_1\mu_2}(\bm{q},\bm{q}_1,\bm{q}_2) \\ &\times S_{\mu_1\nu_1}(q_1,t)S_{\mu_2\nu_2}(q_2,t)\mathcal{Y}^\beta_{\nu \nu_1\nu_2}(\bm{q},\bm{q}_1,\bm{q}_2)^*,\nonumber
\end{align}
where the vertices $  \mathcal{Y}^\alpha_{\mu \mu_1\mu_2}(\bm{q},\bm{q}_1,\bm{q}_2) $ are smooth functions of the control parameters,

\begin{align}\label{eq:vertices}
&\mathcal{Y}^\alpha_{\mu \mu_1\mu_2}(\bm{q},\bm{q}_1,\bm{q}_2) = \frac{n_0}{L^4} \sum_{\kappa} v_{\mu-\kappa}^* \nonumber\\
&\times \left[ b^\alpha(\bm{q}_1 \cdot \bm{q}/q,Q_{\kappa-\mu_2}) c_{\kappa-\mu_2,\mu_1}(q_1) + (1 \leftrightarrow 2)\right].
\end{align}

Here, the direct correlation function $ c_{\mu \nu}(q) $ is defined via the generalized Ornstein-Zernike equation \cite{Hansen:Theory_of_Simple_Liquids,Henderson1992,Lang2010,Lang2012},
\begin{equation}\label{eq:OZ}
\mathbf{S}^{-1} = \frac{n_0}{L^2} \left[ \mathbf{v} - \mathbf{c}  \right],
\end{equation}
and the matrix $ \left[\mathbf{v}\right]_{\mu \nu} = v_{\mu -\nu} $, with $ v(z) = n(z)^{-1} $ corresponds to the local volume. For given control parameters $ \mathbf{S}(q) $, $ \mathbf{c}(q) $ and $ n(z) $, the equations of motion, Eqs.~(\ref{eq:eom1}), (\ref{eq:eom2}), and the mode-coupling functional $ \mathcal{F}^{\alpha \beta}_{\mu \nu} \left[\mathbf{S}(t);q \right] $ define a closed set of integro-differential equations 
with unique solution $\mathbf{S}(q,t) $ that also fulfills all the mathematical constraints  of a correlation function~\cite{Lang_2013}.

Finding a numerical solution of the above equations of motion is, however, a non-trivial task. To stabilize the numerical schemes the introduction of an effective memory kernel $ \mathbf{M}(t) $ has been found to be very useful. To define $ \mathbf{M}(t) $ we employ the Laplace transformation,
\begin{equation}\label{key}
\text{LT}\left\{ \mathbf{A}(t) \right\}(z) = \hat{\mathbf{A}}(z) := {\rm i} \int_0^\infty \mathbf{A}(t) e^{{\rm i} zt} \text{d}t,
\end{equation}
and rewrite Eqs.~(\ref{eq:eom1}) and (\ref{eq:eom2}),

\begin{align}
\hat{\mathbf{S}}(z) &= - [z\mathbf{S}^{-1}+\mathbf{S}^{-1}\hat{\mathbf{K}}(z)\mathbf{S}^{-1}]^{-1},\label{eq:struc1}\\
\hat{\bm{\mathcal{K}}}(z) &= - [z\bm{\mathcal{J}}^{-1}+ {\rm i} \bm{\mathcal{D}}^{-1} +   \hat{\bm{\mathcal{M}}}(z)]^{-1}.\label{eq:struc2}
\end{align}
This allows us to define the effective memory kernel $\hat{{\mathbf{M}}}(z)$ implicitly  via,
\begin{equation}\label{eq:def_meff}
\hat{\mathbf{K}}(z) = -\left[z \mathbf{J}^{-1} + {\rm i} \mathbf{{D}}^{-1} + \hat{\mathbf{M}}(z)  \right]^{-1},
\end{equation}
where the effective matrices   $ \mathbf{J} = \mathcal{C}\{\bm{\mathcal{J}}\}$ and $ \mathbf{J} \mathbf{{D}}^{-1} \mathbf{J}=  \mathcal{C}\{ \bm{\mathcal{J}}\bm{\mathcal{D}}^{-1}  \bm{\mathcal{J}}\}  \succeq 0  $ have been obtained by comparing the high-frequency behavior  of the current kernel $\hat{\mathbf{K}}(z)$.
In the time domain the equation of motion in terms of the effective memory kernel reduces to the standard harmonic oscillator equation with retarded friction
\begin{align}\label{eq:eomS}
\mathbf{J}^{-1}\ddot{\mathbf{S}}(t) + \mathbf{{D}}^{-1}\dot{{\mathbf{S}}}(t) + {\mathbf{S}}(t) \mathbf{S}^{-1}  + \int_{0}^{t} \mathbf{M}(t-t') \dot{\mathbf{S}}(t') \text{d}t' = 0.
\end{align}

\subsection{Diagonal approximation}

To solve Eq.~(\ref{eq:def_meff}) for the effective memory kernel, we rely on the diagonal approximation.  We assume that off-diagonal terms can be discarded in the mode-coupling functional, $ \mathcal{F}^{\alpha \beta}_{\mu \nu}[\mathbf{S}(t),q] = \mathcal{F}^{\alpha}_{\mu}[\mathbf{S}(t),q] \delta_{\alpha \beta} \delta_{\mu \nu} $, in the structure factor, $ S_{\mu \nu}(q) = S_{\mu}(q) \delta_{\mu \nu} $, and the direct correlation function, $ c_{\mu}(q) = 
c_{\mu}(q) \delta_{\mu \nu} $. Consistent with the Ornstein-Zernike equation, Eq.~\eqref{eq:OZ}, we thus set $ v_\mu = 0,\, \forall \mu \neq 0 $ 
and $\mathcal{J}_{\mu \nu}^{\alpha \beta}(q) =  \mathcal{J}_\mu^\alpha(q)  \delta_{\mu \nu}  \delta_{\alpha \beta}$ with   $\mathcal{J}_\mu^\alpha(q)= v_\text{th}^2  $. Furthermore, we assume that the instantaneous damping  $\mathcal{D}^{\alpha\beta}_{\mu\nu}(q) = \delta_{\alpha\beta} \delta_{\mu\nu} \mathcal{D}^\alpha_\mu(q) $ is also diagonal in mode and channel indices.   
As a consequence of the diagonal approximation the coupling of the intermediate scattering functions $ S_\mu(q,t) $ for different wave numbers $ q $ and mode indices $ \mu $ arises purely on the level of the mode-coupling functional.   \gj{It should be noted that the diagonal approximation is a  technical approximation to obtain numerical results for the full time dependence which becomes exact in the planar and bulk limits \cite{limits_cmct}. It has been successfully applied to study the critical packing fraction and the non-ergodicity parameter for confined systems, where it has been compared to the solution without diagonal approximation \cite{Lang2010D} and to computer simulations \cite{Mandal2014,Mandal2017} with good agreement.  }

With this approximation, the left-hand side of Eq.~(\ref{eq:def_meff}) can be rewritten to find,
\begin{align}\label{key}
 &\frac{q^2 }{z v_\text{th}^{-2}+{\rm i}{\mathcal{D}^\parallel_\mu(q)}^{-1}+\hat{\mathcal{M}}^\parallel_\mu(q,z)}\nonumber+\frac{Q_\mu^2 }{z v_\text{th}^{-2}+{\rm i}{\mathcal{D}^\perp_\mu(q)}^{-1}+\hat{\mathcal{M}}^\perp_\mu(q,z)}\\
&= \frac{1}{ z{J_\mu(q)}^{-1}  + {\rm i} {{D}_\mu(q)}^{-1} +  \hat{{M}}_\mu(q,z)  } ,
\end{align}
which can be transformed into an integro-differential equation for the effective memory kernel in the  time domain. For the sake of simplicity we will restrict ourselves to the case of $\mathcal{D}_\mu^\parallel(q) =\mathcal{D}_\mu^\perp(q)=: D_0 $ and thus find,
\begin{align}\label{eq:eomeff}
\dot{M}_\mu&(q,t) + v_\text{th}^2 D_0^{-1}{M}_\mu(q,t) + v_\text{th}^4\int \alpha_\mu(q,t-t') M_\mu(q,t')\text{d}t'=\nonumber\\
& v _\text{th}^4\dot{\beta}_\mu(q,t) + v _\text{th}^6D_0^{-1}{\beta}_\mu(q,t) + \nonumber \\
& v_\text{th}^4J_\mu(q)^{-1}\int_{0}^{t}\mathcal{M}_\mu^{\parallel}(q,t-t')\mathcal{M}_\mu^{\perp}(q,t') \text{d}t'. 
\end{align}

Here, we have used that within the approximations introduced above 
\begin{align}
J_\mu(q) &= (q^2 + Q_\mu^2) v_{\text{th}}^2, \\ 
D_\mu(q) &= (q^2  + Q_\mu^2)D_0, 
\end{align}
and abbreviated
\begin{align}
\alpha_\mu(q,t) &=J_\mu(q)^{-1} ( Q_\mu^2 \mathcal{M}_\mu^\parallel(t)+q^2\mathcal{M}_\mu^\perp(t)),\label{eq:alpha}\\
\beta_\mu(q,t) &=J_\mu(q)^{-2} ( q^2 \mathcal{M}_\mu^\parallel(t)+Q_\mu^2\mathcal{M}_\mu^\perp(t)).\label{eq:beta}
\end{align}
The initial value for the effective memory kernel is generally given by $ \mathbf{J}\mathbf{M}(t=0)\mathbf{J} = - \mathbf{J} \mathbf{D}^{-1} \mathbf{J} \mathbf{D}^{-1} \mathbf{J} +  \mathcal{C}\left\{  \bm{\mathcal{J}}\bm{\mathcal{D}}^{-1}  \bm{\mathcal{J}} \bm{\mathcal{D}}^{-1}  \bm{\mathcal{J}}  \right\} +  \mathcal{C}\left\{  \bm{\mathcal{J}}\bm{\mathcal{M}}(t=0)  \bm{\mathcal{J}} \right\} $. In our case this reduces to,
\begin{equation}\label{eq:initial}
{M}_\mu(q,t=0) = v_\text{th}^4\beta_\mu(q,t=0).
\end{equation}

\section{Numerical Solution of the MCT Equations}
\label{sec:asymptotic_analysis}

In this section we investigate the intermediate scattering function (ISF) $ S_\mu(q,t) $ and the associated frequency-dependent dynamic susceptibility
\begin{align}
 \chi_\mu^{\prime \prime}(q,\omega) := \omega \int_0^\infty S_\mu(q,t) \cos(\omega t) \diff t, 
\end{align}
within the diagonal approximation.
The static input functions, namely the structure factor $ S_\mu(q) $, the direct correlation function $ c_\mu(q) $ and the density profile $ n(z) $ are calculated for monodisperse hard spheres using fundamental measure theory (FMT) and the Ornstein-Zernike equation with the Percus-Yevick closure, as described in Refs.~\cite{Lang2010,Lang2012,Petersen_2019}. The instantaneous damping terms  are  set to $D_0 =  0.1 \,v_{\text{th}}\sigma $  in this work. The particle diameter $\sigma$ sets the unit of length and $\sigma/v_{\text{th}}$ the unit of time.
For the results   the equations of motion, Eqs. (\ref{eq:eomS}) and (\ref{eq:eomeff}), are solved numerically by combining techniques presented in Ref.~\citen{Sperl2000,Gruber2016,Gruber2019,Jung:JStatMech:2020,MC:2020}. Appendix~\ref{sec:numerics} provides details on the numerical discretization scheme which are important to achieve the desired numerical accuracy. For the numerical Fourier transform for the susceptibilities we rely on the modified Filon-Tuck algorithm \cite{Abramowitz19070,Tuck1967}.

\subsection{Dynamics close to the glass transition}

\begin{figure}
	\includegraphics[scale=0.9]{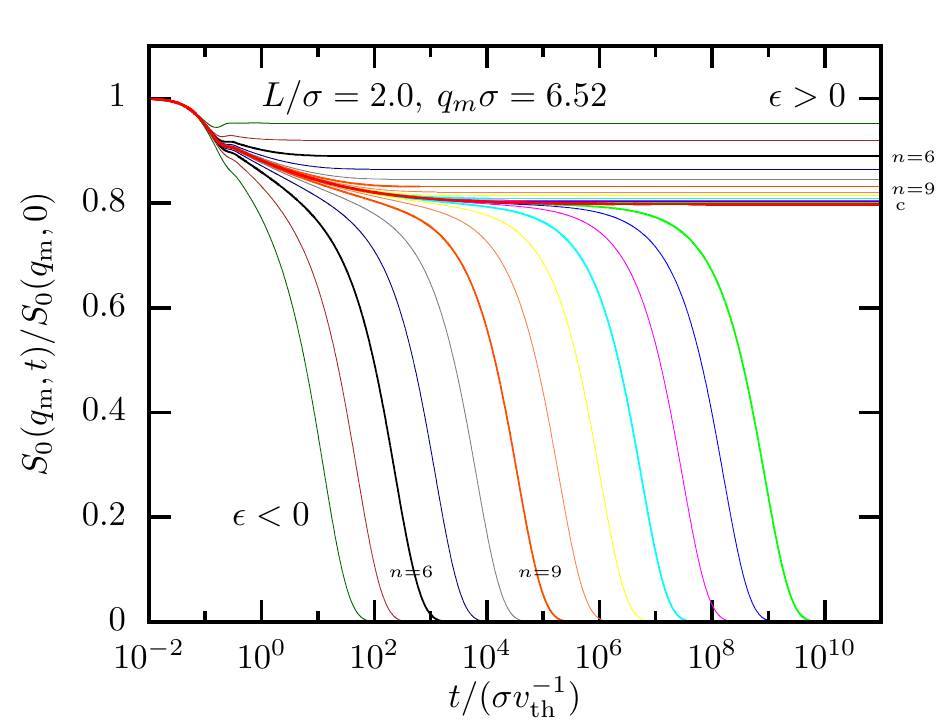}
		\caption{Normalized coherent intermediate scattering function $ S_0(q_\text{m},t)/S_0(q_\text{m},0) $ for accessible width $ L=2.0 \sigma$,  for  wavenumber $ q_\text{m} \sigma = 6.52 $ corresponding to  the first sharp diffraction peak in the structure factor. In the control parameter  $ \epsilon= (\varphi - \varphi_\text{c})/\varphi_\text{c}=\pm 10^{-n/3} $, $ n\in \mathbb{N} $ increases from left to right for $ \epsilon<0 $ and from top to bottom for $ \epsilon > 0 $. The critical correlator for $ \varphi=\varphi_\text{c} $  (or $ \epsilon = 0 $) is displayed as thick line and labeled as ``c''. 
               } 
		\label{fig:coherent}
\end{figure}

\begin{figure}
	\includegraphics[scale=0.9]{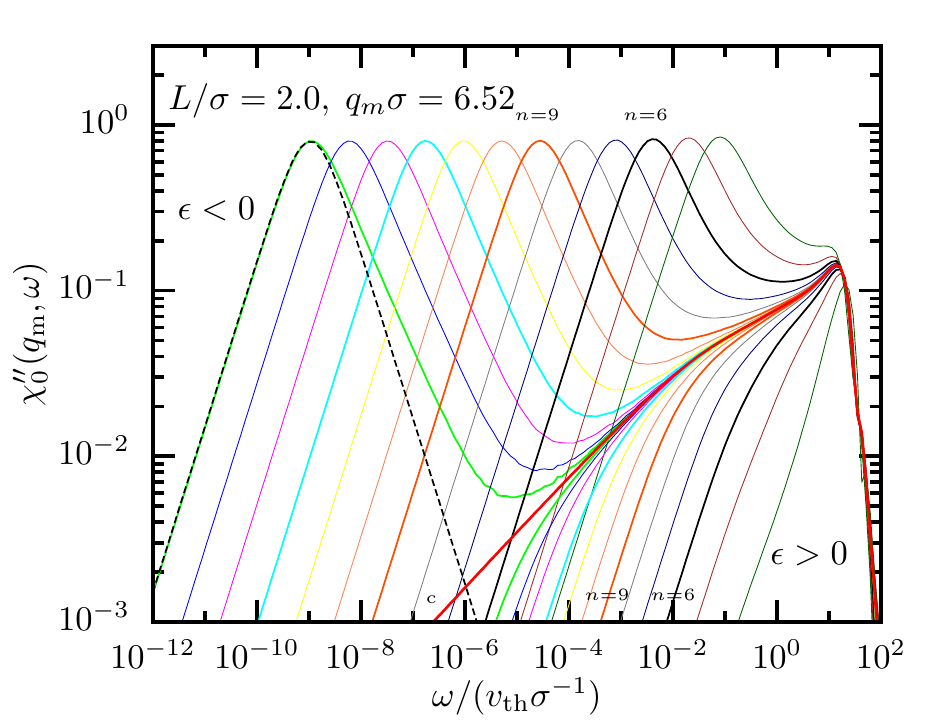}
	\caption{Frequency-dependent susceptibility $ \chi_0^{\prime \prime}(q_\text{m},\omega)$ for the same parameters  as in  Fig.~\ref{fig:coherent}. The dashed, black line shows a Debye peak, $ \chi^{\prime \prime}_{\text{D}}(\omega) = 2 \chi_{\text{max}} \omega \tau_{ \text{\tiny D}}/ \left[ 1+ (\omega \tau_{ \text{\tiny D}} )^2  \right] $ ($ \chi_{\text{max}} = 0.8 $, $ \tau_{ \text{\tiny D}} = 9.4\cdot 10^{8} {\sigma v_\text{th}^{-1} }$) {for comparison}.}
	\label{fig:coherent_FT}
\end{figure}

We start the analysis for an accessible width $ L=2.0 \sigma $. The (normalized) intermediate scattering functions close to the glass transition is displayed in Fig.~\ref{fig:coherent}.
As it is known from simple bulk liquids, the {dynamics in the vicinity of the} glass transition manifests itself within MCT via  {an algebraic}  decay to an {extended}  plateau.   In the liquid regime ($ \epsilon = (\varphi - \varphi_\text{c})/\varphi_\text{c} < 0 $) {this plateau is} followed by a structural relaxation on a time scale, that diverges as the critical packing fraction is approached. Importantly, for any fixed finite time $ t $, the intermediate scattering function $ S_\mu(q,t) $ varies smoothly with the packing fraction. Above the critical packing fraction ($ \epsilon > 0 $), the structure is not able to fully relax anymore and ergodicity breaking is observed, characterized by a non-zero value for the long-time limit, $ F_\mu(q) :=  S_\mu(q,t\to \infty). $ 
The two-step relaxation scenario in the supercooled regime can also be observed in the susceptibility spectrum (see Fig.~\ref{fig:coherent_FT}). {For frequencies much smaller than the microscopic ones, the susceptibilities display a power-law increase $\propto \omega^a$ reflecting the critical decay towards the plateau in the time domain. On the liquid side a second power law $ \propto \omega^{-b}$ emerges at even lower frequencies corresponding to the von Schweidler law as the initial part of the terminal structural relaxation. Both power-law processes are connected by a pronounced minimum which is by orders of magnitude enhanced relative to a  trivial superposition of Debye peaks. The low-frequency peak is known as $\alpha$-peak and appears stretched on the high-frequency flank whereas it behaves regularly at its low-frequency flank.}
   A comparison of these spectra to the ones of simple bulk liquids \cite{Franosch1997} shows that in the case of comparatively large channel widths no qualitative differences {are} observed.

\begin{figure}
	\includegraphics[scale=0.9]{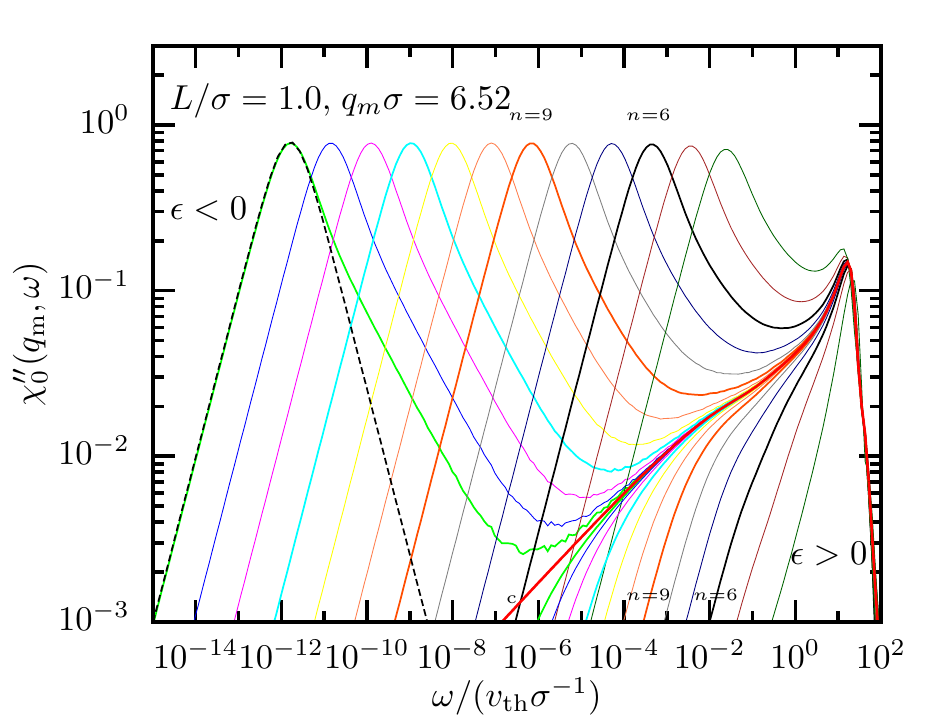}
	\caption{Frequency-dependent susceptibility $ \chi_0^{\prime \prime}(q_\text{m},\omega)$ for channel width $ L = 1.0\, {\sigma} $. The parameters for the Debye peak (dashed black line) are $ \chi_{\text{max}} = 0.79 $, $ \tau_{ \text{\tiny D}} = 6.2\cdot 10^{11} {\sigma v_\text{th}^{-1}} $.}
	\label{fig:coherent_FT_1}
\end{figure}

The situation is {rather} different for a channel width $ L=1.0 \sigma$ as shown in Fig.~\ref{fig:coherent_FT_1}. While the two-step relaxation scenario is still observable there is a distinct kink {in the high-frequency flank of the $\alpha$-peak of } the dynamic susceptibility. A similar feature was recently observed in Ref.~\citen{Jung:JStatMech:2020} for a Bosse-Krieger model with two decay channels. Concluding from this model, it seems natural that the reason for the emergence of the kink is an asymmetry of the two decay channels,  parallel and perpendicular to {the} wall, since this asymmetry increases for decreasing wall separation. We will discuss this observation further in section \ref{sec:confinement}. 

In the following we analyze in detail the two relaxation processes: the critical decay and the von Schweidler law.

\subsection{$ \beta $-scaling regime}

The critical spectra shown in the last {sub}section indicate that the $ \beta $-scaling regime exists also in case of strong confinement. This is not surprising since the asymptotic analysis for MCT with multiple relaxation channels in Ref.~\citen{Jung:JStatMech:2020} showed the existence of a well defined $ \beta $-scaling equation under very moderate assumptions, which are all fulfilled here (i.e. discontinuous transition in all modes). This enables us to perform an asymptotic analysis of the MCT equations in confinement similar to the one performed for simple bulk liquids \cite{Franosch1997}. For this we first determine the critical packing fraction {$\varphi_\text{c}$} using the standard iteration for the nonergodicity parameter $ F_\mu(q) $ (see e.g. Ref.~\citen{Lang2010}). Using Eq.~(19) in Ref.~\citen{Jung:JStatMech:2020} we can then determine the Frobenius-Perron eigenvector and thus {G\"otze's exponent parameter} $ \lambda $ which encodes the critical exponents $a$ and $b$. The important quantities that were determined from the described analysis can be found in Table~\ref{tab:asymptotic}.

\renewcommand{\arraystretch}{1.2}
\begin{table}
	\centering \begin{tabular}{ccccccccc} 
		$ L {/\sigma} $ & $ \varphi_\text{c} $  & $\tilde{\lambda}$ & $\lambda$ & $ a $  & $ b $ & $ B $ &  $ t_0{/ \sigma v_\text{th}^{-1}} $& $ t_\sigma^\prime {/ \sigma v_\text{th}^{-1}} $ \\
		$ 1.0 $ & $ 0.4497 $ & $ 0.816 $ & $ 0.795 $ & $0.282 $ & $ 0.484 $ & $ 1.16 $ & $ 0.022 $ & $ 7.07\cdot 10^{15} $ \\
		\myrowcolour
		$ 1.25 $ & $ 0.4029 $ & $0.631 $ & $0.629$ & $ 0.354 $ & $ 0.761 $&$ 0.41 $& $ 0.035 $&  $ 3.32\cdot 10^{11} $\\
		$ 1.5 $ & $ 0.3817 $ & $ 0.630 $ & $ 0.629 $   & $0.354 $ & $ 0.761 $&$ 0.41 $&$ 0.056 $ & $ 8.68\cdot 10^{11} $\\
		\myrowcolour
		$ 1.75 $& $ 0.4352 $ &  $ 0.676 $ & $ 0.672 $   & $0.338 $ & $ 0.688 $& $ 0.57 $& $ 0.030 $&$ 1.27\cdot 10^{13} $\\ 
		$ 2.0 $& $ 0.4495 $ & $ 0.671 $ &  $ 0.668 $ & $ 0.340 $ & $ 0.694 $& $  0.57 $& $ 0.028 $&$ 2.13\cdot 10^{12} $\\ 
	\end{tabular} 
	\caption{Critical packing fractions $ \varphi_\text{c} $ and asymptotic coefficients for the confinements lengths considered in this work. The universal coefficient $ B $ was interpolated from the data in Ref.~\citen{Gotze_1990}. The time scale $ t_0 $ is determined from matching the critical law to the numerical results for $ \mu = 0 $ and $ q= 6.52 $. The time scale $ t_\sigma^\prime $ for determined for $ \varphi = \varphi - 10^{-7} $  (see App.~\ref{ap:asymptotic} for definitions and further details). }
	\label{tab:asymptotic}
\end{table} 
\renewcommand{\arraystretch}{1.0}

\begin{figure}
	\includegraphics[scale=1]{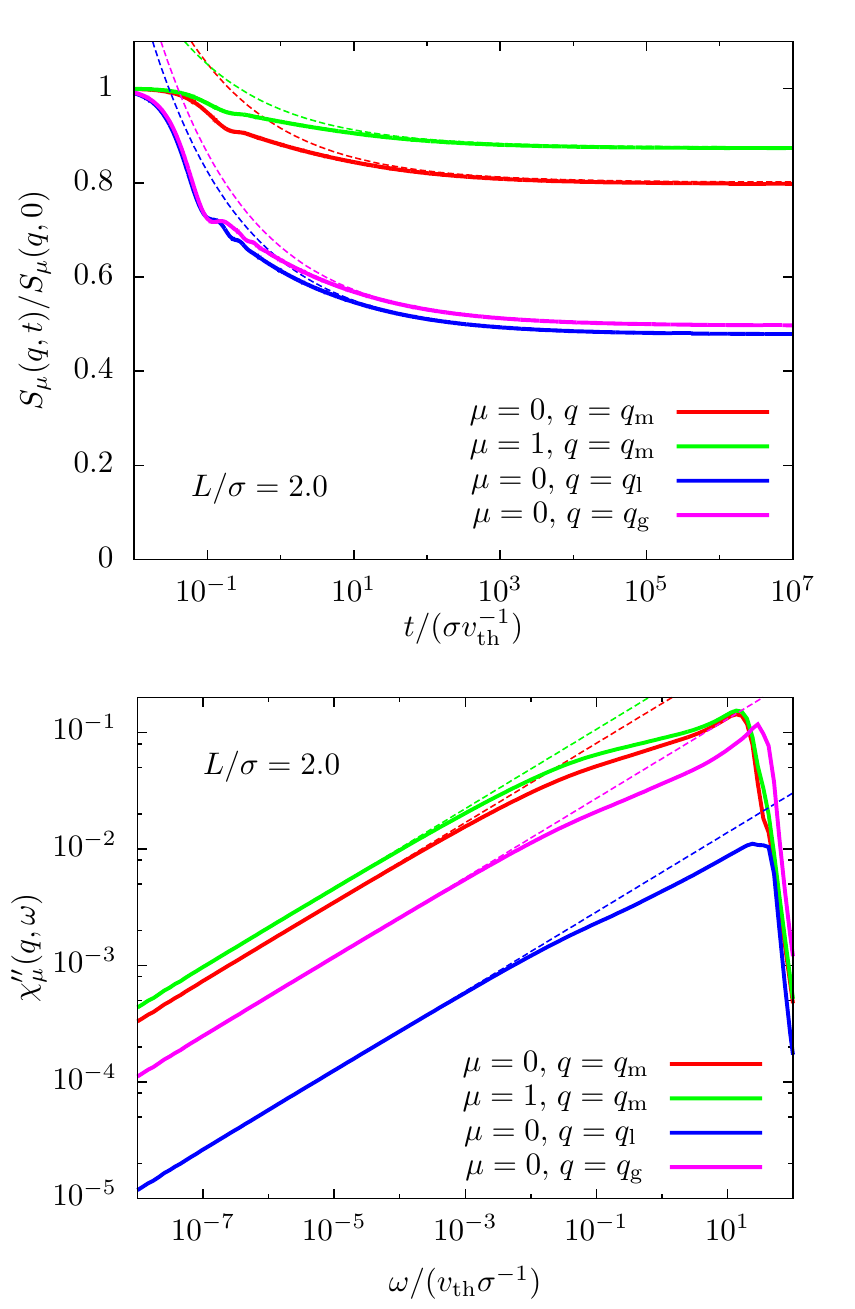}
	\caption{Critical law highlighted in both the normalized coherent scattering function (upper panel) and the frequency-dependent susceptibility (lower panel) for accessible width $ L =2.0 \sigma$ and $ \varphi = \varphi_\text{c} $. Shown are different modes $ \mu $ and wavenumbers slightly below ($ q_\text{l} {\sigma} = 3.42 $), directly at ($ q_\text{m}{\sigma} = 6.52 $) and slightly above ($ q_\text{g}{\sigma} = 9.63 $) the first maximum in the {structure factor $S_0(q,0)$}.  The asymptotes in the upper and lower panel 
correspond to Eqs.~(\ref{eq:crit_asymptote}) and (\ref{eq:crit_asymptote_freq}), respectively. The time scale $ t_0 = 0.028 {\sigma v_\text{th}^{-1}}$ was determined by matching {to} the asymptotic solution of $ S_0(q_\text{m},t) $.  The parameters for the asymptotic analysis are summarized in Table~\ref{tab:asymptotic}. }  
 
	\label{fig:crit}
\end{figure}

\begin{figure}
	\includegraphics[scale=1]{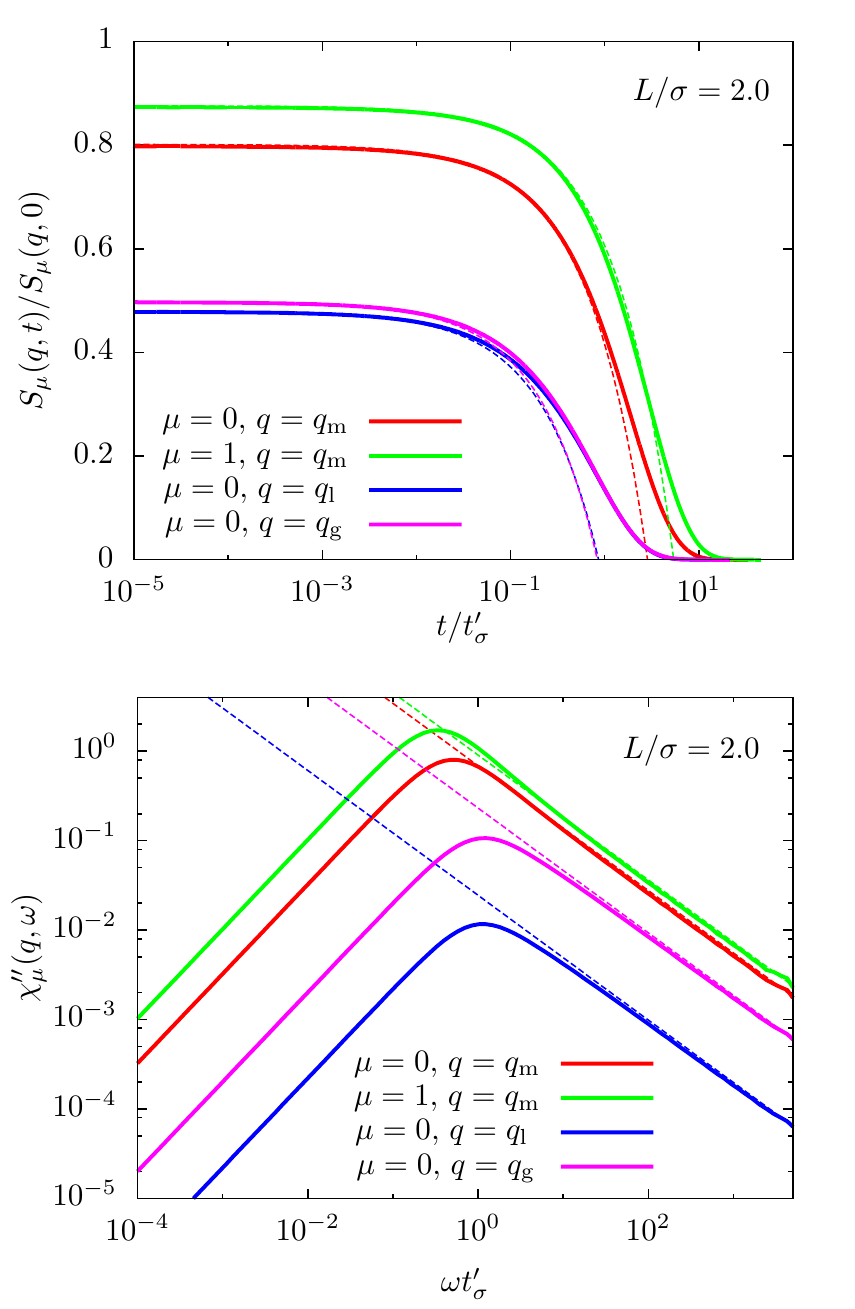}
	\caption{ Same as Fig.~\ref{fig:crit} for the normalized intermediate scattering function (upper panel) and frequency-dependent susceptibility (lower panel) in the $\alpha$-relaxation regime for accessible width $ L=2.0\sigma $ and $ \varphi = \varphi_\text{c} - 10^{-7} $. The asymptotes in the upper and lower panel correspond to Eqs.~(\ref{eq:alpha_asymptote}) and (\ref{eq:alpha_asymptote_freq}), respectively.  The parameters for the asymptotic analysis are summarized in Table~\ref{tab:asymptotic}.}
	\label{fig:schweidler}
\end{figure}

The critical decay of the intermediate scattering function and the frequency-dependent susceptibility spectra are shown in Fig.~\ref{fig:crit} for different mode indices $ \mu $ and wave vectors $ q $. For times {$ t \gtrsim 10^3 { \sigma v_\text{th}^{-1} } $ }  (frequencies {$ \omega \lesssim 10^{-3} { v_\text{th} \sigma^{-1} } $}) there is no visible discrepancy between the asymptotic power laws and the numerical solution of the MCT equations. The decay from the plateau in the intermediate scattering function and the corresponding asymptotic power law for a packing fraction slightly below the critical value $ \varphi_\text{c} $ is then highlighted in Fig.~\ref{fig:schweidler} (upper panel).  The lower panel demonstrates the validity of the von Schweidler law. In these figures no qualitative difference to the critical dynamics in bulk systems can be observed. \gj{It should, however, be emphasized that these results crucially depend on the correct fundamental constant $ \lambda $ which is different from the constant $ \tilde{\lambda} $ one would find without splitting of the relaxation channels such as in bulk liquids (see Table~\ref{tab:asymptotic} and App.\@~\ref{ap:asymptotic})}. Depending on the dissimilarity of the two decay channels this can lead to significant corrections that need to be taken into account to match the numerical solution and the asymptotic power laws. 

The most important discrepancy between bulk and confined liquids are the observed {power-law exponents} which significantly depend on the channel width. This quantitative impact of confinement on the asymptotic scaling laws will be analyzed in the next {sub}section.

\subsection{Effect of confinement}
\label{sec:confinement}

The effect of confinement on the {nonequilibrium-state} diagram has been studied in detail in Refs.~\citen{Lang2010, Lang2012,Mandal2014, Mandal2017a}. In these works a multiple reentrant glass transition was found which can also be observed in computer simulations \cite{Mandal2014, Mandal2017a}. The reason for this behavior can be {rationalized qualitatively} with the competition of two length scales: the (average) particle diameter $ \bar{\sigma} $ and the wall separation $ H $. If the ratio of wall separation and average particle diameter $ n=H/\bar{\sigma} $ is an integer, $ n \in \mathbb{N} $, there are naturally $ n $ different layers in the system which enable a relatively large longitudinal diffusion. For half-integer $ n $, however, there are particles between the layers which significantly slows down the dynamics (incommensurate packing). This non-monotonic dependence on the channel width then becomes apparent in the structure factor, the critical packing fraction and the diffusion coefficient as was shown in Refs.~\citen{Lang2010, Lang2012,Mandal2014, Mandal2017a}.  Here, we will study its impact on the critical {power-law exponents}.

\begin{figure}
	\includegraphics[scale=1]{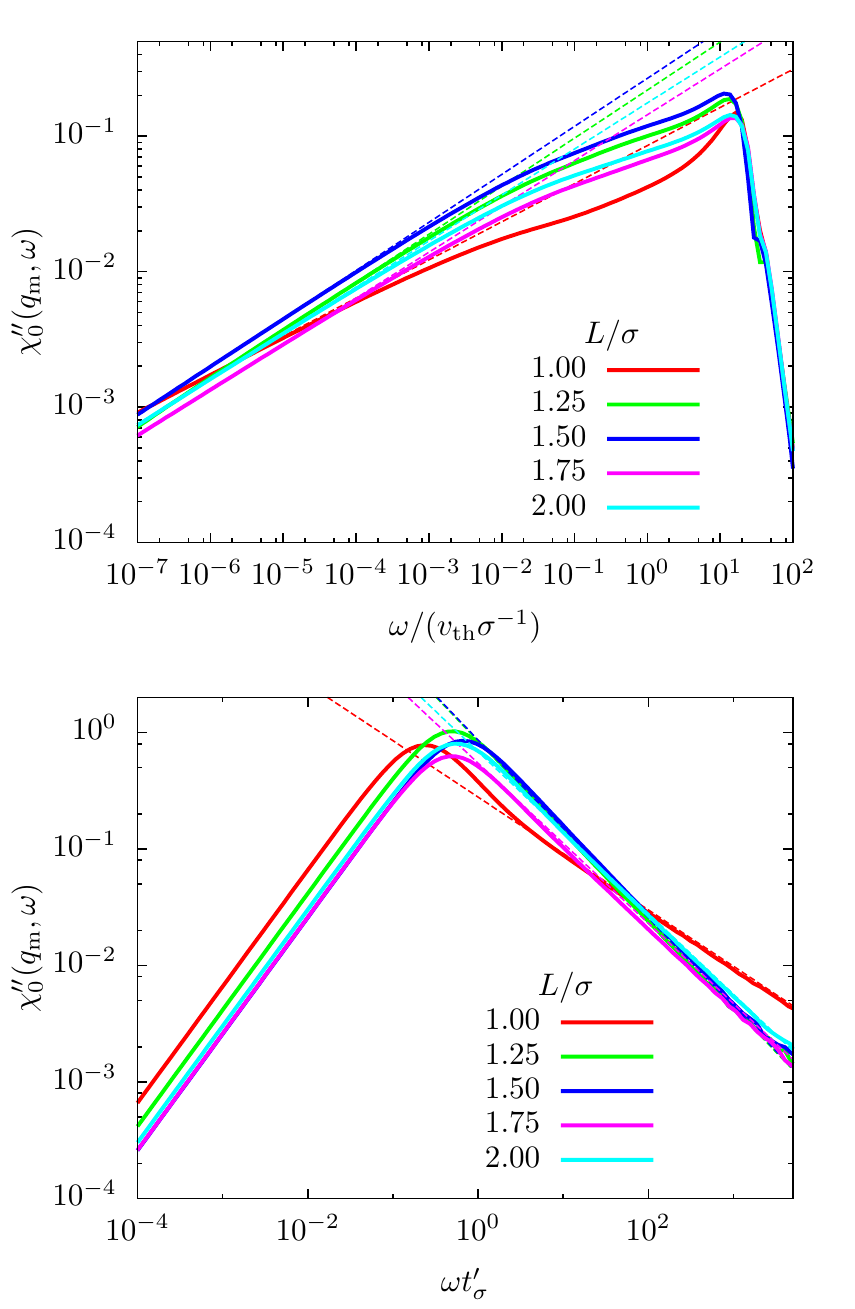}
	\caption{Critical law (upper panel, $ \varphi = \varphi_\text{c} $) and von Schweidler law (lower panel, $ \varphi = \varphi_\text{c} - 10^{-7} $) for the frequency-dependent susceptibility for different accessible widths $ L $. The parameters for the asymptotic analysis are summarized in Table~\ref{tab:asymptotic}.}
	\label{fig:crit_H}
\end{figure}

\begin{figure}
	\includegraphics[scale=0.93]{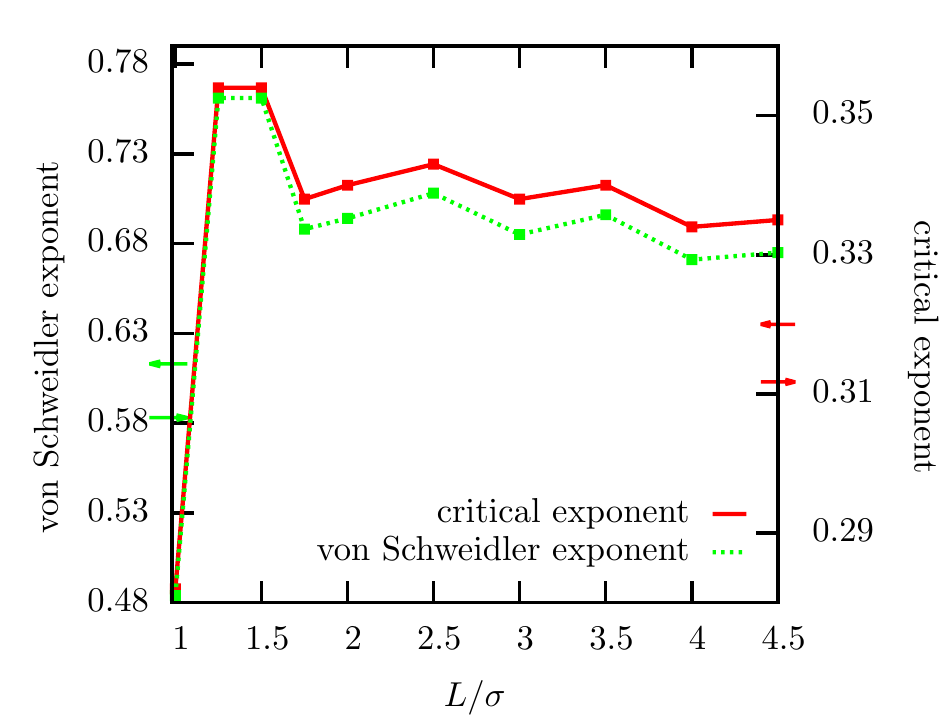}
	\caption{Dependence of the critical exponents on the accessible width $L$. The exponents were determined from the asymptotic analysis as described in Ref.~\citen{Jung:JStatMech:2020}. \gj{The arrows indicate the value of the exponents in the bulk limit for hard spheres (arrows pointing to the right \cite{Franosch1997}) and hard discs (arrows pointing to the left \cite{2D_MCT}). }}
		\label{fig:exp}
\end{figure}

{The} critical decay and the von Schweidler law {(lower panel)} for the frequency-dependent susceptibility for different accessible widths $L$ { are displayed in Fig.~\ref{fig:crit_H}}. The most apparent difference between the curves is the significantly different power{-}law exponent for $L=1.0 \sigma$ (strong confinement). To investigate this in more detail, the exponents are plotted {vs.} accessible width in Fig.~\ref{fig:exp}. {Strikingly}, similar to the static quantities, also this plot shows a non-monotonic behavior of the critical exponent $ a $ and the von Schweidler exponent $ b $ with wall separation. This means that moderate confinement $ L>1.25 \sigma$ reduces the stretching of the correlation functions relative to the bulk system (an unstretched exponential would correspond to $ b\rightarrow1 $) but a further decrease of the accessible width then leads to stronger stretching. This shows that the effect of confinement has indeed a non-trivial impact on all features of {glass-forming} liquids. \gj{Interestingly, the convergence of the exponents to the bulk limit for hard spheres is very slow. We explain this with the inhomogeneous density profiles, which significantly alter the glass transition and are strongly pronounced even for large channels $ L > 4.5\sigma. $ Furthermore, the critical packing fraction of the glass transition increases significantly with $ L $ (from $ \varphi_c(L=2.5\sigma) = 0.437 $ to $ \varphi_c(L=4.5\sigma) = 0.469 $) which additionally amplifies the layering.  }

\begin{figure}
	\includegraphics[scale=1]{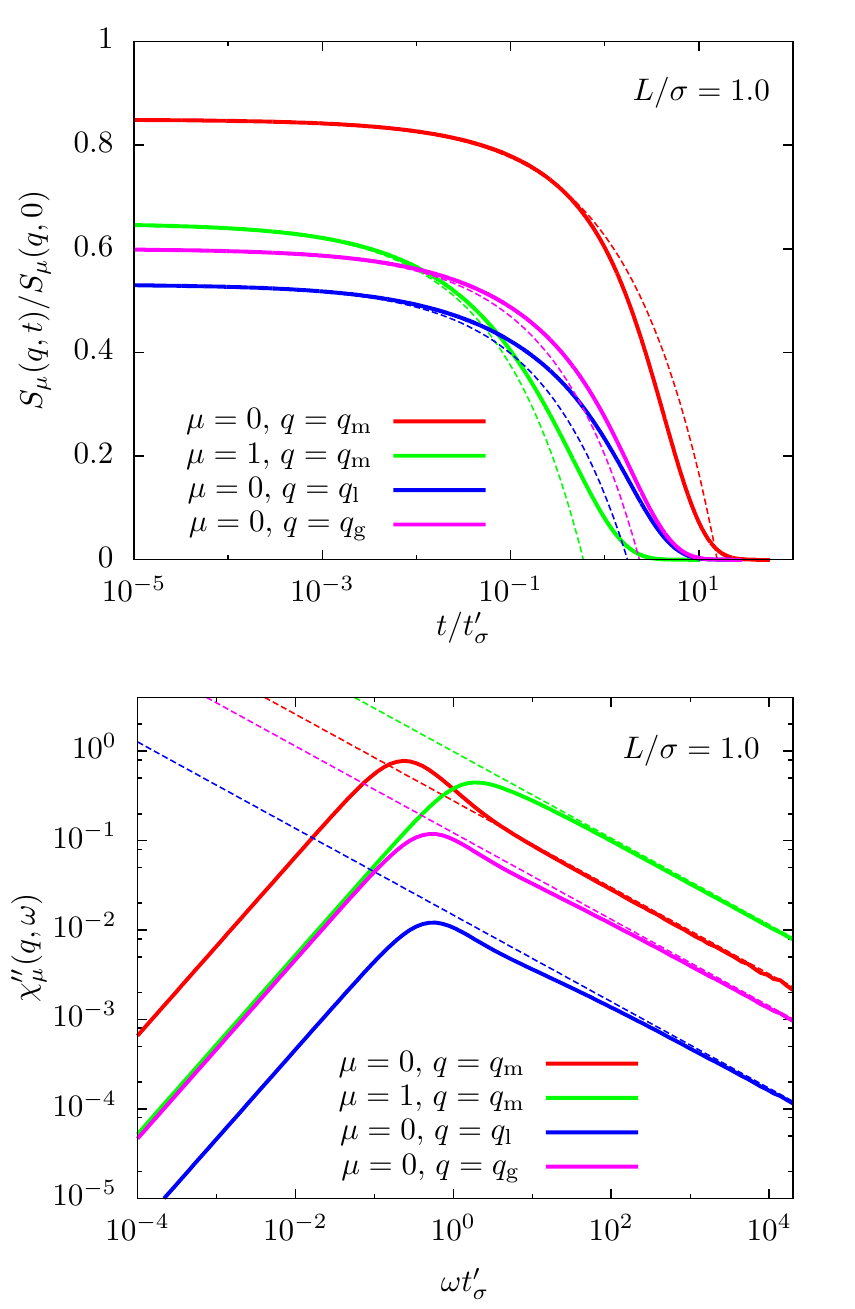}
	\caption{Normalized intermediate scattering function (upper panel) and frequency-dependent susceptibility (lower panel) in the $\alpha$-relaxation regime for channel width $ L=1.0 \sigma $. Shown are different modes $ \mu $ and wave vectors slightly below ($ q_\text{l} {\sigma} = 3.42 $), directly at ($ q_\text{m} {\sigma} = 6.52 $) and slight above ($ q_\text{g} {\sigma}= 9.63 $) 
the first maximum in the {structure factor $S_0(q)$}. The parameters for the asymptotic analysis are summarized in Table~\ref{tab:asymptotic}. }
	\label{fig:schweidler1}
\end{figure}

We now come back to the ``kink'' observed in the low-frequency susceptibility spectrum for $ L=1.0 \sigma$ (highlighted in Fig.~\ref{fig:schweidler1}). A similar kink in the low frequency-spectrum was before observed as Cole-Cole peak \cite{Sperl:PhysRevE.74.011503}, however, there the kink emerges for frequencies larger than the von Schweidler law. We rationalize the kink therefore in the same way as was discussed in Ref.~\citen{Jung:JStatMech:2020}. It can be observed in Fig.~\ref{fig:schweidler1} (upper panel) that the green curve ($ \mu=1 $) decays faster than the other curves due to a large ratio $ H_\mu(q)/F_\mu(q) $, where $  H_\mu(q) $ is the Frobenius-Perron eigenvector of the critical expansion. Additionally, the two memory kernels that define the relaxation of the red curve ($ \mu =0 $) are very different since the parallel component couples stronger to itself then to other modes while the perpendicular component does not couple at all to itself and has a strong coupling to other modes. We therefore have a similar situation as discussed in Ref.~\citen{Jung:JStatMech:2020} where we introduced a toy model with two very different decay channels. We thus draw a similar conclusion: When the higher modes decay faster (represented here by the green curve), we can observe multiple low-frequency peaks in the susceptibility spectrum, each corresponding to a different relaxation channel. Obviously, in the present case the dissimilarity between the decay times of the different modes is quite small which means that the two peaks strongly overlap. This leads to the observed ``kink''.

\section{Comparison to event-driven molecular dynamics simulations}
\label{sec:comparison_edmd}

\begin{figure}
	\includegraphics[scale=0.9]{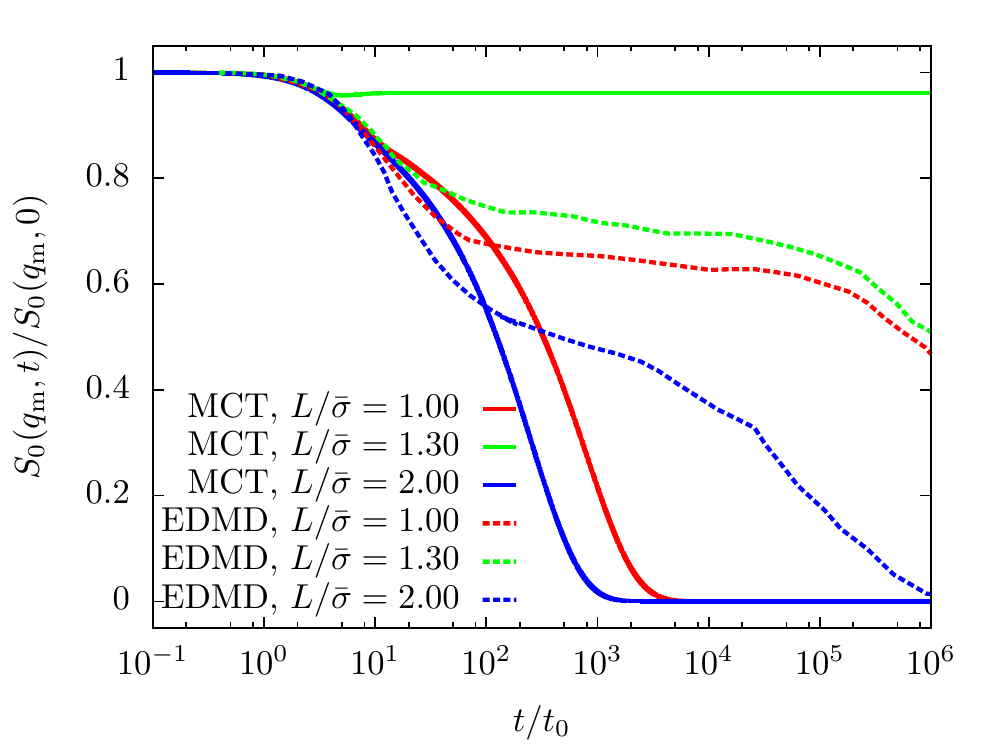}
	\caption{Comparison of MCT results for the coherent scattering function at volume fraction $ \varphi = 0.42 $ with event-driven molecular dynamics (EDMD) simulations at $ \varphi = 0.52 $. {Simulation data are extracted from Ref.~\citen{Mandal2017a}}. The simulations use a polydisperse mixture ($ s=15\% $) and $ \bar{\sigma} $ denotes the average particle diameter. } 
	\label{fig:coherent_sim}
\end{figure}

Event-driven molecular dynamics (EDMD) simulations \cite{Alder1957a,Rapaport1980,Bannerman2011} enable the exact integration of {equations of motion} for particles with hard-sphere interactions. 
{Here we compare our theoretical results to simulation results that we have extracted from Ref.~\citen{Mandal2017a}. There the authors collected data for polydisperse hard spheres in confined geometry as in the current set-up}. 
  The introduction of polydispersity is necessary to suppress crystallization.   The full numerical solution of the MCT equations presented in this manuscript now {allows} us to directly compare theoretical and simulation results for the dynamical properties of confined hard-sphere glasses (see Fig.~\ref{fig:coherent_sim}).

On the one hand, significant quantitative differences between theory and simulations are expected because the polydispersity significantly reduces the effect of layering in the system and the differences between commensurate and incommensurate packing. This can already be seen in the phase diagram of the polydisperse hard-sphere system (see Fig.~6 (a) in Ref.~\citen{Mandal2014}) and thus also reduces the differences in the intermediate scattering functions for various accessible widths $ L. $ Additionally, there is no ideal glass transition in simulation (and experiments) of hard spheres because eventually the structure will fully relax, as is already known from simple bulk liquids.

On the other hand, there are several features that are shared by theory and simulations. Most prominent is the fact that also in this case structural relaxation is slower for the system with $ L=1.3 \bar{\sigma}$ than for $ L=1.0 \bar{\sigma}$ despite the general trend that confinement slows down the dynamics. This leads to a clear order of the curves which is the same for both theory and simulations. Furthermore, both theory and simulations exhibit a pronounced two-step relaxation scenario that was discussed in detail in the previous section. 

 It is also noteworthy that the dependence of the von Schweidler exponent on accessible width $ L $ as predicted by MCT is in slight contradiction with event-driven simulations. In Ref.~\citen{Scheidler_2002,Mandal2017a} simulation results indicated that stretching could be strongest for incommensurate packing, which contradicts the conclusion drawn for MCT in the last section.  One possible reason for the discrepancy could be the different control-parameter distances $ \epsilon = (\varphi - \varphi_\text{c}(L))/\varphi_\text{c}(L) $ since the simulations for different accessible width $ L $ were all performed at constant packing fraction $ \varphi $ while the results reported for the theory were determined for $  \varphi_\text{c}(L) $ and thus $ \epsilon=0 $. If $ \left | \epsilon \right | $ becomes too large this will have an impact on the simulation results.

\section{Summary and conclusion}
\label{sec:conclusion}

In this manuscript we have presented a numerical solution of the MCT equations of motion in slab geometry.
This enabled us to perform a deep analysis of the dynamics of confined hard-sphere glasses.
We {have } found that there are no qualitative differences between the $ \beta $-scaling regime in systems with moderate confinement and bulk systems.
Only for very small accessible width $ L {\lesssim} 1.25 \sigma$ a clear feature of the parallel relaxation scenario induced by the confinement {has been} observed in the form of a kink in the low-frequency susceptibility spectrum.

Additionally, we {have applied} an asymptotic analysis for MCTs with multiple relaxation channels to investigate the effect of confinement on the two asymptotic power laws: the critical decay and the von Schweidler law.  
We {have} observed that, similar to the non-monotonic dependence of the critical packing fraction, also the {power-law exponents} depend non-trivially on the channel width.
For moderate confinement we {have} found that the stretching of the intermediate scattering function is decreased (corresponding to an increase of the von Schweidler exponent) while it is significantly increased for strong confinement, indicating stronger heterogeneous dynamics in the slit. \gj{Importantly, mode-coupling theory also predicts a very slow convergence of the critical exponents to the bulk limit for hard spheres. }

The numerical results have also been compared to event-driven molecular dynamics simulations of confined, polydisperse hard spheres. {Both} theory and simulations exhibit a clear {non-monotonic dependence of their dynamic properties on the channel width. {Yet, a  comparison beyond identifying the relevant  trends  is unfeasible} due to the different polydispersities and the known discrepancies of the ideal transition as predicted by MCT to the dynamic crossover in simulations and experiments. 

This work opens up the possibility of a throughout analysis of various aspects of confined fluids with mode-coupling theory. Possible future projects are the incorporation of Brownian dynamics \cite{Schrack:2020} and the study of single particle properties like self-intermediate scattering functions \cite{Lang2014b} or mean-square displacements. Importantly, the latter would enable a significantly better comparison with computer simulations. This can for example facilitate the search for a similar kink as observed in the MCT solution of the low-frequency susceptibility spectrum. Additionally, it will be interesting to look for signatures of parallel relaxation in experimental measurements of dielectric spectra of molecules or confined particles. \gj{Furthermore, using static quantities from computer simulations for extreme confinement could enable the cMCT analysis of the dimensional crossover to the planar bulk limit for hard discs \cite{2d_limit_structural,fluids_extreme}. Currently this is not possible since the iterative techniques to determine the static input functions do not converge in extreme confinement.}

\section*{Acknowledgments} 
We gratefully acknowledge inspiring discussions with Suvendu Mandal. We thank Matthias Sperl for drawing our attention to the Cole-Cole peak. 
This work has been supported by the Austrian Science Fund (FWF): I 2887.

\appendix

\section{Details on the numerical discretization scheme}
\label{sec:numerics}

The correlation functions appearing in this manuscript depend on a mode index, a wavevector and time. In the following, we will describe how these dependences are handled to obtain a numerical integration scheme for the MCT equations of motion. The parameters used in this manuscript are summarized in Table~\ref{tab:parameters}.

\renewcommand{\arraystretch}{1.1}
\begin{table}
	\centering \begin{tabular}{cc|cc|cc} 
		\multicolumn{2}{c}{FMT}  & \multicolumn{2}{c}{OZ + PY} & \multicolumn{2}{c}{MCT} \\
		$ \text{d}z/\sigma $ & \hphantom{11} $ 0.001 $ \hphantom{11} & $ d_z/\sigma $ & 0.02 & $ N_q $ &  \hphantom{11} $ 30 $  \hphantom{11} \\
		\myrowcolour
		$ n_z $ & $ 8192 $ & $ N_q $ & 1024 & $ M $ & $ 5 $  \\
		$ \mathcal{F}^\text{ex} $ & WBII & $ r_\text{min}/\sigma $ & 0.0001   & $\Delta q \sigma $ & $ 1.0345 $ \\
		\myrowcolour
		&  &  $ r_\text{max}/\sigma $ &\hphantom{11} 10000  \hphantom{11} & $N_t $ & 8192\\ 
		&  &  &   & $ D $ & 100\\ 
		\myrowcolour
		&  &  &   &$ D_1 $ & 22   \\ 
		&  &  &  & $ \Delta t_0/(\sigma v_\text{th}^{-1}) $ &  \hphantom{11} $ 10^{-9} $  \hphantom{11} \\
		\myrowcolour
		&  &  &  & $ \epsilon_s $ & $ 10^{-8} $ \\
	\end{tabular} 
	\caption{Summary of the parameters used for creating the static input using fundamental measure theory (FMT) and the Ornstein-Zernike equation with the Percus-Yevick closure (OZ+PY) as well as the numerical integration of the mode-coupling equations of motion (\ref{eq:eomS}) and (\ref{eq:eomeff}). \textbf{FMT}) The parameters $ \text{d}z $ and $ n_z $ were used for the spatial linear discretization. The approximate excess free energy functional used is the White-Bear mark II (WBII) functional \cite{fmt:Rosenfeld1989,fmt:Roth2010}. \textbf{OZ+PY}) The parameter $ \text{d}z $ was used for the spatial linear discretization in z-direction with $ n_z = L/\text{d}z $. The r-direction is logarithmically discretized in the range $ [r_\text{min} , r_\text{max}] $ with $ N_q $ grid points. Details in Refs.~\citen{Lang2010D,Petersen_2019}. \textbf{MCT}) The equations are temporally integrated using $ D $ decimation steps with $ N_t $ time steps each and an initial step size $ \text{d}t_0 $.}
	\label{tab:parameters}
\end{table} 
\renewcommand{\arraystretch}{1.0}

\subsection{Mode index}

The mode index is already discrete due to the finite channel width. However, for the numerical solution we have to introduce a cutoff $ |\mu| \leq M $, chosen such that the neglected orders do not have a significant impact on the main modes anymore. Empirically we found $ M=5 $ to be a reasonable cutoff for $ L\lesssim 2{\sigma} $. For the results in Fig.~\ref{fig:exp} we use $ M=10 $ for $ L \lesssim 3.5 \sigma  $ and $ M=15 $ otherwise.

\subsection{Wavevector}

As done in Ref.~\citen{Lang2010} we first introduce the thermodynamic limit of Eq.~(\ref{eq:MCT_functional}) by replacing,
\begin{align}\label{eq:asymptote}
\lim\limits_{N,A\rightarrow \infty} &\frac{1}{N} \sum_{\substack{\bm{q}_1,\\\bm{q}_2=\bm{q}-\bm{q}_1}} (...) = \frac{1}{n_0 (2\pi)^2} \int_0^\infty \text{d}q_1 \nonumber \\
\cdot &\int_{|q-q_1|}^{q+q_1} \text{d}q_2 \frac{4q_1q_2}{\sqrt{4q_1^2q_2^2 - (q^2 + q_1^2 - q_2^2)^2}}(...).
\end{align}
Here, $ A $ is the surface area of the box with volume $ V=HA. $ The $ q $-dependence is then discretized to $ q = q_0 + m \Delta q  $, with $ m = 0,...,N_q-1 $. The integral is evaluated using a modified trapezoidal rule in which the value of the function to be integrated is not taken in the middle of two grid points, $q_{m}=m\Delta q$ and $q_{(m+1)}$, but at $q_m+q_0$, with $ q_0=0.303\Delta q $.
This provides the best discrete description of the Jacobian of the transformation to bipolar coordinates which allows to write the right term in Eq.~\eqref{eq:asymptote}~\cite{Bayer2007,MC:2020}. In this work we chose $ N_q = 30 $ and $ \Delta q\sigma = 1.0345 $ such that $ q_\text{max} \sigma = 30 $. This choice allows us the solve the equations of motion with the necessary accuracy in the dynamical correlation functions while the numerical error due to the $ q- $discretization remains small.

\subsection{Time}

To find a numerical solution of the integro-differential equations, {Eqs.}~(\ref{eq:eomS}) and (\ref{eq:eomeff})), we  need to discretize their {time dependence} such that the derived integration schemes are stable for about 15-20 orders of magnitude in time. To achieve this goal, we apply similar schemes as previously proposed in the MCT literature, like decimation \cite{Sperl2000,Gruber2019,Jung:JStatMech:2020}, which shall not be repeated here. In the following, we will list some important remarks that are necessary to stabilize the numerical discretization scheme.

\begin{itemize}
	\item As discussed in Ref.~\citen{Jung:JStatMech:2020}, Appendix~B, we find that taking the {time derivative} on both sides of Eq.~(\ref{eq:eomeff}) and discretize the obtained second order integro-differential equation yields a significantly more stable numerical scheme for longer times. Starting from  decimation step $ D_1 $ {we} therefore integrated the memory kernel with this second-order equation which has the same form as Eq.~(\ref{eq:eomS}). The applied discretization scheme is thus the same as for the integration of the scattering function.
	\item We symmetrize the discretized convolution integrals in the same way as described in Ref.~\citen{Jung:JStatMech:2020}, Appendix~B.
	\item The instantaneous contributions to the memory kernels, $ \mathbf{D}^{-1} $ and $ \bm{ \mathcal{D}}^{-1} $, are explicitly included to the values of  $ \mathbf{M}(t=0) $ and $ \bm{ \mathcal{M}}(t=0) $, respectively. We thus obtain, 
	\begin{align}
	\tilde{M}_\mu(q,t=0) &= \beta_\mu(q,t=0)  + {D}_\mu(q)^{-1}/\Delta t,\\
	\tilde{\mathcal{M}}_\mu^\alpha(q,t=0) &= \mathcal{F}_\mu^\alpha[\mathbf{S}(t=0);q]  + D_0^{-1}/\Delta t.
	\end{align}
	
	This yields a more compact numerical integrator. The integration scheme using moments (see Refs.~\cite{Sperl2000,Gruber2019}) ensures that these contributions are handled correctly in the decimation steps.
\end{itemize}

 Taking all this into account we arrive at the following integration scheme for the coherent scattering function (not including the explicit dependence on $ q $ and $ \mu $),
\begin{align}\label{eq:intS1}
A_S	S_i  &=   2.5 S_{i-1} - 2 S_{i-2} + 0.5 S_{i-3} \nonumber\\
	& \hspace*{-0.5cm} + \frac{\Delta t^2 J}{4} \left( 2 \,\text{d}\tilde{M}_1 S_{i-1} + 2\tilde{M}_{i} S_{0} - \tilde{M}_{i-\bar{i}} S_{\bar{i}} - \tilde{M}_{\bar{i}} S_{i-\bar{i}}  \right)  \nonumber\\
	&\hspace*{-0.5cm}- \frac{\Delta t^2J}{2} \sum_{j=1}^{\bar{i}} \text{d}S_j ( \tilde{M}_{i-j+1} - \tilde{M}_{i-j})   \nonumber\\
		&\hspace*{-0.5cm}- \frac{\Delta t^2J}{2}   \sum_{j=2}^{\bar{i}} \text{d}\tilde{M}_j ( S_{i-j+1} - S_{i-j})  \nonumber\\
	&\hspace*{-0.5cm}- \frac{\Delta t^2J}{4}\begin{cases}
	\text{d}S_{i-\bar{i}} ( \tilde{M}_{\bar{i}+1} - \tilde{M}_{\bar{i}})  & \text{if }\bar{i} \neq i - \bar{i}\\
	0 & \text{otherwise}
	\end{cases}\nonumber\\
		&\hspace*{-0.5cm}- \frac{\Delta t^2J}{4}\begin{cases}
		\text{d}\tilde{M}_{i-\bar{i}} ( S_{\bar{i}+1} - S_{\bar{i}})  & \text{if }\bar{i} \neq i - \bar{i}\\
		0 & \text{otherwise}
		\end{cases}\nonumber,\\
	A_S &=  1 + \frac{\Delta t^2 J}{2} \left( S_0^{-1} + d\tilde{M}_1 \right),
\end{align}
with $ S_i = S(i\Delta t) $, $ \text{d}S_i = \Delta t^{-1} \int_{(i-1)\Delta t }^{i\Delta t} \text{d}t' S(t') $ and similarly $ M_i $, $ \text{d}M_i $. We also introduced $ \bar{i}=\lfloor i/2  \rfloor  $. The brackets $ \lfloor j \rfloor $ denote the largest integer less or equal $ j $. Before decimation step $ D_\text{1} $ the effective memory kernel is integrated via,
\begin{align}\label{eq:intM1}
A_{M_1}	M_i  &=   4/3 M_{i-1} - 1/3 M_{i-2} - \frac{\Delta t^2 v_\text{th}^4}{3}  \alpha[\text{d}\mathcal{M}_1] M_{i-1} \nonumber  \\
& \hspace*{-0.5cm} +  v_\text{th}^4\left( \beta[\mathcal{M}_i] - 4/3  \beta[\mathcal{M}_{i-1}]+ 1/3  \beta[\mathcal{M}_{i-2}] \right) \nonumber\\
&\hspace*{-0.5cm}- \frac{\Delta t^2v_\text{th}^4}{3}   \sum_{j=1}^{i-\bar{i}} \text{d}M_j ( \alpha[\mathcal{M}_{i-j+1}] + \alpha[\mathcal{M}_{i-j}])\nonumber \\
&   \hspace*{-0.5cm}- \frac{\Delta t^2v_\text{th}^4}{3}   \sum_{j=2}^{\bar{i}} \alpha[\text{d}\mathcal{M}_j] ( M_{i-j+1} + M_{i-j})  \nonumber\\
&   \hspace*{-0.5cm}+ \frac{\Delta t^2v_\text{th}^4J^{-1}}{3}   \sum_{j=1}^{\bar{i}} \text{d}\mathcal{M}^\parallel_j ( \mathcal{M}^\perp_{i-j+1} + \mathcal{M}^\perp_{i-j})  \nonumber\\
&   \hspace*{-0.5cm}+ \frac{\Delta t^2v_\text{th}^4J^{-1}}{3}   \sum_{j=1}^{i-\bar{i}} \text{d}\mathcal{M}^\perp_j ( \mathcal{M}^\parallel_{i-j+1} + \mathcal{M}^\parallel_{i-j})  \nonumber\\
A_{M_1} &=  1 + \frac{\Delta t^2v_\text{th}^4}{3} \alpha[\text{d}\mathcal{M}_1],
\end{align} 
where we defined $ \alpha[\mathcal{B}] = J_\mu(q)^{-1} ( Q_\mu^2 \mathcal{B}_\mu^\parallel(t)+q^2\mathcal{B}_\mu^\perp(t))  $ and $ \beta[ \mathcal{B} ] = J_\mu(q)^{-2} ( q^2 \mathcal{B}_\mu^\parallel(t)+Q_\mu^2\mathcal{B}_\mu^\perp(t)) $. As discussed before, after decimation step $ D_1 $ we change the integration algorithm and use,
\begin{align}\label{eq:intM2}
A_{M_1}	M_i  &=   2.5 M_{i-1} -2 M_{i-2}+0.5 M_{i-2} \nonumber \\
& \hspace*{-0.5cm}+ \frac{\Delta t^2v_\text{th}^4}{4} \left( 2\alpha[\text{d}\mathcal{M}_1] M_{i-1} -  \alpha[\mathcal{M}_{i-\bar{i}}] M_{\bar{i}} - \alpha[\mathcal{M}_{\bar{i}}] M_{i-\bar{i}}  \right)   \nonumber  \\
& \hspace*{-0.5cm} +  v_\text{th}^4\left( \beta[\mathcal{M}_i] - 2.5  \beta[\mathcal{M}_{i-1}]+ 2  \beta[\mathcal{M}_{i-2}] - 0.5\beta[\mathcal{M}_{i-3}] \right) \nonumber\\
&\hspace*{-0.5cm}- \frac{\Delta t^2v_\text{th}^4}{2}   \sum_{j=1}^{\bar{t}} \text{d}M_j ( \alpha[\mathcal{M}_{i-j+1}] - \alpha[\mathcal{M}_{i-j}])\nonumber \\
&   \hspace*{-0.5cm}- \frac{\Delta t^2v_\text{th}^4}{2}   \sum_{j=2}^{\bar{t}} \alpha[\text{d}\mathcal{M}_j] ( M_{i-j+1} - M_{i-j})  \nonumber\\
&\hspace*{-0.5cm}- \frac{\Delta t^2v_\text{th}^4}{4}\begin{cases}
\text{d}M_{i-\bar{i}} ( \alpha[\mathcal{M}_{\bar{i}+1}] - \alpha[\mathcal{M}_{\bar{i}}])   & \text{if }\bar{i} \neq i - \bar{i}\\
0 & \text{otherwise}
\end{cases}\nonumber\\
&\hspace*{-0.5cm}- \frac{\Delta t^2v_\text{th}^4}{4}\begin{cases}
\alpha[\text{d}\mathcal{M}_{i-\bar{i}}] ( M_{\bar{i}+1} - M_{\bar{i}} )  & \text{if }\bar{i} \neq i - \bar{i}\\
0 & \text{otherwise}
\end{cases}\nonumber\\
& \hspace*{-0.5cm}+ \frac{\Delta t^2v_\text{th}^4J^{-1}}{4} \left(    \mathcal{M}^\parallel_{i-\bar{i}} \mathcal{M}^\perp_{\bar{i}} + \mathcal{M}^\perp_{i-\bar{i}} \mathcal{M}^\parallel_{\bar{i}}  \right)   \nonumber  \\
&\hspace*{-0.5cm}+ \frac{\Delta t^2v_\text{th}^4J^{-1}}{2}   \sum_{j=1}^{\bar{t}} \text{d}\mathcal{M}^\parallel_j ( \mathcal{M}^\perp_{i-j+1} - \mathcal{M}^\perp_{i-j})\nonumber \\
&   \hspace*{-0.5cm}+ \frac{\Delta t^2v_\text{th}^4J^{-1}}{2}  \sum_{j=1}^{\bar{t}} \text{d}\mathcal{M}^\perp_j ( \mathcal{M}^\parallel_{i-j+1} - \mathcal{M}^\parallel_{i-j})  \nonumber\\
&\hspace*{-0.5cm}+ \frac{\Delta t^2v_\text{th}^4J^{-1}}{4} \begin{cases}
\text{d}\mathcal{M}^\parallel_{i-\bar{i}} ( \mathcal{M}^\perp_{\bar{i}+1} - \mathcal{M}^\perp_{\bar{i}})   & \text{if }\bar{i} \neq i - \bar{i}\\
0 & \text{otherwise}
\end{cases}\nonumber\\
&\hspace*{-0.5cm}+\frac{\Delta t^2v_\text{th}^4J^{-1}}{4} \begin{cases}
\text{d}\mathcal{M}^\perp_{i-\bar{i}} ( \mathcal{M}^\parallel_{\bar{i}+1} - \mathcal{M}^\parallel_{\bar{i}} )  & \text{if }\bar{i} \neq i - \bar{i}\\
0 & \text{otherwise}
\end{cases},\nonumber\\
A_{M_1} &=  1 + \frac{\Delta t^2v_\text{th}^4}{2} \alpha[\text{d}\mathcal{M}_1].
\end{align} 
In each time step, $ S_i $ is initialized by setting $S_i = S_{i-1} $. With this the memory kernels $ \mathcal{M}^\alpha_\mu(q) $ are calculated using Eq.~(\ref{eq:MCT_functional}) in the thermodynamic limit as described in Eq.~\eqref{eq:asymptote}. $ S_i $ is then determined self-consistently by solving Eqs.~\eqref{eq:intS1} and \eqref{eq:intM1}/\eqref{eq:intM2} until the convergence reaches an accuracy of $ \max\limits_{q,\mu}| S_{i,\mu}^n(q) - S_{i,\mu}^{n-1}(q) | <  \epsilon_s $ in the $ n $-th iteration step as described in Refs.~\cite{Haussmann1990,Sperl2000}.

\section{Asymptotic expansion}
\label{ap:asymptotic}

\newcommand{\disphat}[2][3mu]{\hat{#2\mkern#1}\mkern-#1}

\gj{In Ref.~\cite{Jung:JStatMech:2020} an asymptotic analysis of mode-coupling-theory equations with multiple relaxation channels has been presented. The reference proves the validity of the $ \beta $-scaling equation and derives relations for the critical exponents that characterize the slowing down at the glass transition. In the following, we recapitulate the most important relations and describe how they can be applied to the mode-coupling equation in confined geometry.}

\gj{Starting from the equations for structural relaxation, Eqs.~(\ref{eq:struc1}) and (\ref{eq:struc2}), with negligible contributions of $z \bm{\mathcal{J}} + {\rm i} \bm{\mathcal{D}}^{-1}$, we perform an asymptotic expansion using the ansatz $ \mathbf{S}(q,t) - \mathbf{F}_\text{c}(q) = \sqrt{\left| \sigma \right|} \mathbf{G}^{(1)}(t) + \mathcal{O}(\sigma) $, for a small separation parameter $ \sigma $. (In this appendix, we follow the standard MCT notation and $\sigma$ denotes a separation parameter, not the hard-sphere diameter.)  We thus assume that the correlator is close to its plateau value (i.e. the critical non-ergodicity parameter $  \mathbf{F}_\text{c}(q) $).   It has been shown that close to the glass transition, $ \sigma = C\epsilon $, with constant $ C $ and control parameter $ \epsilon = (\varphi - \varphi_\text{c})/\varphi_\text{c} $. }

\gj{To first order we find the factorization theorem,
\begin{equation}\label{eq:factorization}
\mathbf{G}^{(1)}(q,t) = {\mathbf{H}}(q) {g}(\hat{t}=t/t_\sigma), 
\end{equation}
stating that close to the glass transition on a time scale $ t_\sigma $ all dynamical correlation functions can be rescaled by the critical amplitudes $ {\mathbf{H}}(q) $ to superimpose on a single universal master curve $ {g}(\hat{t}) $.}

\gj{The equation of motion for this master curve is then derived as solubility condition by considering the second order of the expansion~\cite{Jung:JStatMech:2020},
\begin{align}\label{eq:beta_scaling2}
\frac{\text{d}}{\text{d} \hat{t}} ({g} \ast {g}) (\hat{t}) ={\lambda} {g}(\hat{t})^2 + \text{sgn}\, \sigma,
\end{align}
which is the well-known $ \beta- $scaling equation. The exponent parameter 
$ \lambda $ connects the power law exponents for the critical decay, $ a, $ and the von-Schweidler law, $ b $, via \emph{G\"otze's exponent relation},
\begin{equation}\label{eq:Goetze_exp}
\frac{\Gamma(1+b)^2}{\Gamma(1+2b)} = \lambda = \frac{\Gamma(1-a)^2}{\Gamma(1-2a)}.
\end{equation}
The subtleties derived in Ref.~\cite{Jung:JStatMech:2020} are that this $ \beta $-scaling equation is only found via rescaling, which is possible due to its scale invariance. From the asymptotic analysis we obtain,
\begin{equation}\label{eq:lambda}
{\lambda} = \tilde{\lambda}/(1-\Delta),
\end{equation}
with channel asymmetry $ \Delta = 0 $ in case of bulk geometry. The parameters $ \mathbf{H}(q), C, \Delta $ and $ \tilde{\lambda} $ are complicated functions of the mode-coupling functional in confined geometry (and therefore also the static input functions) as well as the critical non-ergodicity parameters (see Ref.~\cite{Jung:JStatMech:2020} for details). }

\gj{The workflow to apply the asymptotic expansion is the following:
\begin{itemize}
	\item Find the critical packing fraction $ \varphi_c $ using binary search based on the asymptotic equation,
	\begin{align}\label{eq:nonergodic1}
	\mathbf{S}(q) - \mathbf{F}(q)  =  [ \mathbf{S}(q)^{-1} + \mathbf{N}(q) ]^{-1}
	,
	\end{align}
	where $  \mathbf{N}(q)= \mathcal{C} \Big\{ \bm{\mathcal{F}}\left[ \mathbf{F}(q);q\right]^{-1} \Big\}^{-1} $. This equation can be readily used as self-consistent iteration scheme to determine the non-ergodicity parameter \cite{Lang2010,Lang2012} and such the ideal glass transition.
	\item Evaluate the critical non-ergodicity parameter, $  \mathbf{F}_\text{c}(q) $, to calculate the mode-coupling functional at the critical point and thus the critical amplitude (using the eigenvalue equations (27)-(29) in Ref.~\cite{Jung:JStatMech:2020}).
	\item Calculate the parameters $ \tilde{\lambda} $  and $ \Delta $ using Eqs.~(38) and (39)  in Ref.~\cite{Jung:JStatMech:2020}. From this determine the power law exponents $ a $ and $ b $ with Eqs.(\ref{eq:lambda}) and (\ref{eq:Goetze_exp}).
	\item For $ \epsilon \neq 0 $, we can also directly calculate $ \sigma $ using the non-ergodicity parameter $ \mathbf{F}_\epsilon(q) $ for $ \varphi = \varphi_\text{c}(1+\epsilon) $ (see Eq.~(37) in Ref.~\cite{Jung:JStatMech:2020}).
\end{itemize}}

\gj{It is important to note that $ \Delta \approx 0 $ and thus $ \tilde{\lambda} \approx \lambda $ implies that the relaxation channels parallel and perpendicular to the walls become very similar. It does, however, not mean that confinement has no influence on the critical exponents, since the expansion is still based on the full mode-coupling functional for confined geometry.}

\gj{The asymptotic dynamics of the correlation functions can be directly extracted from the $ \beta $-scaling equation, similar to bulk liquids (see Refs.~\citen{Franosch1997,Gotze2009}). To summarize:
\begin{itemize}
	\item For times $ t $ much larger than the microscopic ones and $ {t} \ll t_\sigma $ the short-time solution $ g(\hat{t} \ll 1) = \hat{t}^{-a} $ sets $ t_\sigma = t_0 \left| \sigma\right|^{-1/2a} $ and we obtain,
	\begin{align}\label{eq:crit_asymptote}
	\mathbf{S}(q,t) &\simeq \mathbf{F}_\text{c}(q) + \mathbf{{H}}(q) (t/t_\sigma)^{-a} \sqrt{\left|\sigma \right|} ,\\
	\boldsymbol{\chi}''(q,\omega) &\simeq \mathbf{{H}}(q)  \Gamma(1-a)\sin\left(\pi a/2\right)(\omega t_\sigma)^a\sqrt{\left|\sigma \right|} .\label{eq:crit_asymptote_freq}
	\end{align}
	Here, $ \boldsymbol{\chi}''(q,\omega) = \omega \mathbf{S}''(q,\omega) $ is the dynamic susceptibility, determined from the Fourier cosine transform of the correlation function, $ \mathbf{S}''(q,\omega) = \int_0^\infty \cos (\omega t ) \mathbf{S}(q,t) \text{d}t. $
	\item For $ \sigma \geq 0 $ and $ {t} \gg t_\sigma $ the non-ergodicity parameter is given by,
	\begin{equation}\label{eq:static_asymptote}
	\mathbf{F}(q) = \lim\limits_{t\rightarrow \infty }	\mathbf{S}(q,t) \simeq \mathbf{F}_\text{c}(q) + \mathbf{{H}}(q) \sqrt{\frac{\sigma}{1-\lambda}}.	
	\end{equation}
	\item For $ \sigma < 0 $ and $  {t} \gg t_\sigma $ a second power law emerges, $ g(\hat{t} \gg 1 ) = -B\hat{t}^b $, corresponding to the early $ \alpha $-relaxation on a time scale $ t'_\sigma = (t_0/B^{1/b})\left| \sigma \right|^{-\gamma} $, with $ \gamma = 1/2a + 1/2b $. For the correlation function and the dynamic susceptibility we find,
	\begin{align}\label{eq:alpha_asymptote}
	\mathbf{S}(q,t) &\simeq \mathbf{F}_\text{c}(q) - \mathbf{{H}}(q) (t/t'_\sigma)^{b},\\
	\boldsymbol{\chi}''(q,\omega) &\simeq \mathbf{{H}}(q)  \Gamma(1+b)\sin\left(\pi b/2\right)(\omega t'_\sigma)^{-b}.\label{eq:alpha_asymptote_freq}
	\end{align}
\end{itemize}}

\FloatBarrier

\bibliography{library_local}

\end{document}